\documentclass[twocolumn,amsmath,amssymb,prb,superscriptaddress]{revtex4-1}

\usepackage{graphicx}
\usepackage{dcolumn}
\usepackage{bm}
\usepackage{hyperref}
\usepackage{xcolor}

\allowdisplaybreaks
\usepackage{multirow}

\begin{document}

\title{Universal transport dynamics in a quenched tunnel-coupled Luttinger liquid}
\author{F. M. Gambetta}
\affiliation{Dipartimento di Fisica, Universit\`{a} di Genova, Via Dodecaneso 33, I-16146 Genova, Italy}
\affiliation{SPIN-CNR Genova, Via Dodecaneso 33, I-16146 Genova, Italy}
\author{F. Cavaliere}
\affiliation{Dipartimento di Fisica, Universit\`{a} di Genova, Via Dodecaneso 33, I-16146 Genova, Italy}
\affiliation{SPIN-CNR Genova, Via Dodecaneso 33, I-16146 Genova, Italy}
\author{R. Citro}
\affiliation{Dipartimento di Fisica ``E. R. Caianiello'', Universit\`a degli Studi di Salerno, Via Giovanni Paolo II 132, I-84084 Fisciano (Salerno), Italy}
\affiliation{SPIN-CNR Salerno, Via Giovanni Paolo II 132, I-84084 Fisciano (Salerno), Italy}
\author{M. Sassetti}
\affiliation{Dipartimento di Fisica, Universit\`{a} di Genova, Via Dodecaneso 33, I-16146 Genova, Italy}
\affiliation{SPIN-CNR Genova, Via Dodecaneso 33, I-16146 Genova, Italy}

\date{\today}

%*************************************************************************************************
%Abstract
%*************************************************************************************************

\begin{abstract}
The transport dynamics of a quenched Luttinger liquid tunnel-coupled to a fermionic reservoir is investigated. In the transient dynamics, we show that for a sudden quench of the electron interaction {\em universal} power-law decay in time of the tunneling current occurs, ascribed to the presence of entangled compound excitations created by the quench. In sharp contrast to the usual non universal power-law behavior of a zero-temperature non-quenched Luttinger liquid,  the steady state tunneling current is ohmic and can be explained in terms of an effective quench-activated heating of the system. Our study unveils an unconventional dynamics for a quenched Luttinger liquid that could be identified in quenched cold Fermi gases. \\

\noindent PACS number(s): 71.10.Pm, 67.85.Lm, 05.70.Ln, 73.63.-b
\end{abstract}

\maketitle

%*************************************************************************************************
%Introduction
%*************************************************************************************************
\section{Introduction}
Non-equilibrium dynamics of interacting quantum many-body systems~\cite{Polkovnikov:2011,Eisert:2015} has recently gained a lot of interest thanks to the fast experimental progresses in the field of ultracold bosonic and fermionic atomic gases~\cite{Bloch:2008rmp,Bloch:2008s,Bloch:2012}, which have allowed to probe the real time evolution of several quantum many-body systems out-of-equilibrium~\cite{Greiner:2002,Kinoshita:2006,Trotzky:2012,Cheneau:2012,Langen:2015}. In fact, ultracold atomic gases offer key advantages of tuning, with high precision, system parameters, such as the strength of the interaction and dimensionality~\cite{Bloch:2008rmp,Langen:2015,Loftus:2002,Roberts:1998,Greiner:2002b}, making it possible to create and probe local excitations with single-site and real-time resolution \cite{bloch_2011,kantian_2013} or perform transport experiments~\cite{Brantut:2012,Krinner:2015,Husmann:2015,Krinner:2015b}. Very interestingly, when parameters are sweeped in time, cold atoms systems allow to experimentally realize a so called {\em quantum quench} protocol~\cite{Polkovnikov:2011,Eisert:2015} and to study the ensuing quantum dynamics.\\
A widely studied theoretical question has been about the conditions under which one-dimensional (1D) systems eventually thermalize if prepared in the ground state of an initial Hamiltonian $H_i$ and brought out of equilibrium by time evolving them with a final Hamiltonian $H_f$ of similar form but with changed parameters~\cite{Kinoshita:2006,Rigol:2007,Manmana:2007,Rigol:2008,Rigol:2009,Rigol:2009pra}, as the interaction strength~\cite{Cazalilla:2006,Iucci:2009,Uhrig:2009,Cazalilla:2016}. Other studies, instead, have regarded quench protocols with the switching on/off of an external field~\cite{Iucci:2010,Calabrese:2011} or of the coupling between two identical systems~\cite{Perfetto:2006,Perfetto:2010,DallaTorre:2013,Foini:2015}. Concerning the time duration of an interaction quench, both the cases of an abruptly change of the system Hamiltonian (sudden quench)~\cite{Cazalilla:2006,Iucci:2009,Uhrig:2009,Perfetto:2011} and of a slow variation of the latter~\cite{Dora:2011,Bernier:2011,Bernier:2014,Sachdeva:2014}, have been considered.\\

From the point of view of the dynamics following a quantum quench, 1D interacting Fermi systems are promising candidates to realize unusual nonequilibrium steady states. Already in equilibrium they show a peculiar behavior. Indeed, they fall into the Luttinger liquid (LL) universality class, which is characterized by power-law decay of correlation functions
with interaction-dependent exponents~\cite{Voit:1995,Giamarchi:2004,vonDelft:1998,Weiss:1995}. In particular, transport properties show a peculiar power-law suppression with the applied bias of the differential conductance for tunneling through an opaque barrier~\cite{Kane:1992,Matveev:1993,Voit:1993,Bockrath:1999,Guinea:1995,Furusaki:2002}. Furthermore, such a model possesses an infinite number of constants of motion and hence thermalization after a quantum quench is a non-trivial issue~\cite{Rigol:2007,Rigol:2008,Rigol:2009,Cassidy:2011,Polkovnikov:2011,Cazalilla:2016}.  In the wake of the newly performed transport experiments in ultracold atoms~\cite{Brantut:2012,Krinner:2015,Husmann:2015,Krinner:2015b} an increasing interest has grown in studying the interplay between the quench dynamics and the peculiar transport properties of a LL~\cite{Perfetto:2010,Perfetto:2011,Kennes:2013,Kennes:2014,Schiro:2014,Schiro:2015} and recent studies have confirmed the
{\em non-universal} power-law scaling towards an asymptotic steady state in transport properties.

In this work we are interested in the transient dynamics of a interaction-quenched LL tunnel-coupled to a fermionic reservoir.
We find that for any sudden quench protocol of the interaction, with the exception of the one that leads to a noninteracting final state, the tunneling current exhibits a $t^{-2}$ {\em universal} power-law scaling as a function of time after the quench towards its steady state value. The emergence of such {\em universal} behavior is unusual for a LL and can be ascribed to the presence of compound excitations created by the quench~\cite{Calabrese:2006,Bernier:2014,Cazalilla:2016}. 
Our findings are different from previous results obtained in Refs.~\onlinecite{Perfetto:2011,Schiro:2015}, in which only a typical non-universal power-law time-scaling appears. Here, the presence of an external reservoir tunnel-coupled with the bulk of the system is essential for the observability of the universal behavior.
In the long-time limit, the system settles to a steady state with an ohmic tunneling current (and thus with a non-vanishing zero-bias conductance), that could be associated to an effective-temperature effect induced by the quench.   

%*************************************************************************************************
\section{The model}
We consider a spinless LL with periodic boundary conditions, subject to a sudden homogeneous quench of the interaction. The Hamiltonian describing the quenched LL is~\cite{Cazalilla:2006,Iucci:2009}
\begin{equation}
\label{eq:H0_1}
H_{0,\mathrm{LL}}(t)=H_{0,\mathrm{LL}}^{(i)}\theta(-t)+H_{0,\mathrm{LL}}^{(f)}\theta(t),
\end{equation}
where $ \theta(t) $ is the Heaviside step function and, in bosonized form~\cite{Giamarchi:2004,Voit:1995,vonDelft:1998},
\begin{equation}
\label{eq:H0_2}
H_{0,\mathrm{LL}}^{(\nu)}=\sum_{q\neq 0}v_{\nu}|q|\beta_{\nu,q}^{\dagger}\beta_{\nu,q}+\Omega_{\nu}\, .
\end{equation}
Here and in what follows $\hbar=1$, $\nu=i,f$ labels the pre- ($t<0$) or post-quench ($t>0$) state of the LL, $v_{\nu}=v_{\mathrm{F}}/K_{\nu}$ is the plasmon velocity with $ v_F $ the Fermi velocity and $0<K_{\nu}\leq 1$ the LL interaction parameter (with $K_{\nu}<1$ for repulsive interactions and $K_{\nu}=1$ for a non-interacting channel), and $q=2 \pi n/L$ is the momentum, with $n$ an integer and $L$ the length of the LL. Furthermore, $\beta_{\nu,q}$ and $\beta_{\nu,q}^{\dagger}$ are canonical bosonic operators and 
\begin{equation}
\Omega_{\nu}=\frac{Lv_{\mathrm{F}}}{4\pi\alpha^2}\left[\Delta_{f}-\Delta_{i}\right]\delta_{\nu,f}	
\end{equation}
is the ground-state energy mismatch of the post-quench state with respect to the state prior the quench, with
\begin{equation}
\Delta_{\nu}=\frac{1}{K_{\nu}}\left(2-K_{\nu}-\frac{1}{K_{\nu}}\right)\, .
\end{equation}
Here, $\alpha\ll L$ is the shortest-length cutoff. We can interpret $ \Omega_\nu $ as the energy injected into the system during the quench.
In order to maintain the validity of the LL theory, which is a low energy description, the latter must be smaller than the Fermi energy of the system. See Fig.~\ref{fig:fig0}.
\noindent Bosonic operators before and after the quench are connected by the canonical transformation
\begin{equation}
\label{eq:betaf}
\beta_{f,q}\!=\!\frac{1}{2}\!\left(\!\sqrt{\frac{K_i}{K_f}}\!+\!\sqrt{\frac{K_f}{K_i}}\right)\!\beta_{i,q}+\frac{1}{2}\!\left(\!\sqrt{\frac{K_i}{K_f}}\!-\!\sqrt{\frac{K_f}{K_i}}\right)\!\beta_{i,-q}^{\dagger}\, .	
\end{equation}
Note that, since we consider a system in the thermodynamic limit, in Eq.~\eqref{eq:H0_2} we are neglecting the zero mode of the LL.
The LL is tunnel-coupled with a point contact to a non-interacting fermionic reservoir described by the Hamiltonian
\begin{equation}
\label{eq:H0tip}
H_{0,\mathrm{R}}=\sum_{\mathbf{k}}\left[\varepsilon(\mathbf{k})-eV\right]c_{\mathbf{k}}^{\dagger}c_{\mathbf{k}}\, ,
\end{equation}
with $-e$ the electron charge, $V$ the bias w.r.t. the LL and $c_{\mathbf{k}}$ a fermionic operator for electrons in the reservoir with wavevector $\mathbf{k}$. The flow of current through the LL is allowed by the coupling with a second reservoir, which works as a source/drain, placed far away from the point contact and with a tunneling barrier much less opaque. In this configuration the transport properties of the system are dominated by tunneling through the opaque barrier and the presence of the second reservoir is negligible~\cite{WeissBook}. The coupling between reservoir and LL is described via a local tunneling Hamiltonian
\begin{equation}
\label{eq:H_T}
H_{\mathrm{T}}(t)=\mathcal{M}\theta(t-t_0)\sum_{r=\pm}\psi_{r}^{\dagger}(x_0)\psi_{\mathrm{R}}(z_{\mathrm{R}})+\mathrm{H.c.},
\end{equation}
where $t_0>0$ is the time when tunneling is switched on, $x_0$ is the location of the LL where the point contact sits at and $z_{\mathrm{R}}$ is the coordinate in the reservoir from where electrons tunnel. Here, $\mathcal{M}$ is the tunneling amplitude, $r$ an index representing right ($r=+$) and left ($r=-$) branches in the LL, and
\begin{equation}
\label{eq:psir}
\psi_{r}(x)=\frac{F_{r}e^{irq_{\mathrm{F}}x}}{\sqrt{2\pi\alpha}}e^{i r \phi_{r}(x)}
\end{equation}
is the associated fermionic field~\cite{Giamarchi:2004,Voit:1995,vonDelft:1998}, with $F_{r}$ the Klein factor of the branch $ r $, $q_{\mathrm{F}}$ the Fermi wavevector and
\begin{equation}
\label{eq:phir}
\phi_{r}(x)=\sqrt{\frac{2\pi}{L}}\sum_{q\neq 0}\frac{e^{-\alpha |q|/2}}{\sqrt{|q|}}\left[A_{r,q}(x)\beta_{f,rq}+A_{r,q}^{*}(x)\beta^{\dagger}_{f,rq}\right]
\end{equation}
the LL bosonic field. Here, the coefficients
\begin{equation}
\label{eq:Aq}
A_{r,q}(x)=e^{irqx}\left[\theta(q)u_{+}-\theta(-q)u_{-}\right]\, ,
\end{equation}
with $u_{\pm}=(K_{f}^{-1/2}\pm K_{f}^{1/2})/2$, have been introduced. Furthermore,
\begin{equation}
\label{eq:psiT}
\psi_{\mathrm{R}}(z)=\sum_{\mathbf{k}}\Psi_{\mathbf{k}}(z)c_{\mathbf{k}}\,,
\end{equation}
is the fermionic operator for the reservoir, with $\Psi_{\mathbf{k}}(z)$ the electron wavefunctions and $z$ a coordinate in the latter. With the system in thermal equilibrium at $ t\leq0^{-} $, we evaluate the tunneling current in the zero-temperature limit and to the lowest perturbative order in the tunnel coupling, i.e. $ |\mathcal{M}|^2 $, obtaining (see Appendix~\ref{appendix} for details)
\begin{equation}
\label{eq:current}
I(t)=I_0\int_{0}^{t-t_0}\mathrm{d}\tau\,\mathrm{Re}\left[\frac{\sin(eV\tau)}{\tau}f_b(t,t-\tau)\right]\, ,
\end{equation}
where $I_0=4e\mathcal{D}|\mathcal{M}|^2/(\pi\alpha)$, with $ \mathcal{D} $ the density of states of the reservoir. In Eq.~\eqref{eq:current} we have introduced the correlator $ f_{b}(t_1,t_2)=2\pi\alpha\langle\psi_{I}(x_0,t_2)\psi^{\dagger}_{I}(x_0,t_1)\rangle_{i}$, which is independent from the LL branches. Here, the subscript $ I $ stands for interaction picture with respect to $ H_{\mathrm{T}}(t) $ and $ \langle...\rangle_{i} $ represents the average on the initial ground state. We obtain
\begin{equation}
\label{eq:fb}
f_{b}(t_1,t_2)=\mathcal{C}_{-}(t_1-t_2)\mathcal{C}_{+}(t_1-t_2)\mathcal{U}(t_1,t_2)\,,
\end{equation}
with
\begin{align}
\mathcal{C}_{\pm}(t_1-t_2)&=\left[\frac{1}{1\pm i\mathcal{K}(t_1-t_2)}\right]^{\nu_{\pm}}\,,\\
\mathcal{U}(t_1,t_2)&=\left\{\frac{(1+4\mathcal{K}^2t_2^2)(1+4\mathcal{K}^2t_1^2)}{\left[1+\mathcal{K}^2(t_1+t_2)^2\right]^2}\right\}^{\eta}\,,
\end{align}
where $ \mathcal{K}=(\tau_0 K_f)^{-1} $, with $ \tau_0=\alpha/v_F $ the shortest-time cutoff of the theory, and 
\begin{align}
\nu_{\pm}&=\frac{(1+K_{f}^2)(K_{f}\mp K_{i})^2}{8K_{f}^2K_{i}}\, ,\\
\eta&=\frac{(1-K_{f}^2)(K_{f}^2-K_{i}^2)}{8K_{f}^2K_{i}}\, .\label{eq:eta}
\end{align}
In the correlator of Eq.~\eqref{eq:fb} we can distinguish three different contributions. The first one, $ \mathcal{C_{-}}(t_1-t_2) $, is the usual term present in the equal-space Green function of a zero-temperature non-quenched LL, although with the exponent renormalized by the quench, and stems from the bosonic averages $ \langle\beta_{f,q}\beta^{\dagger}_{f,q}\rangle_{i} $. The second term, $ \mathcal{C_{+}}(t_1-t_2) $, comes from the averages $ \langle\beta^{\dagger}_{f,q}\beta_{f,q}\rangle_{i} $, while the third one, $ \mathcal{U}(t_1,t_2) $, arises from the ``anomalous averages'' $ \langle\beta_{f,q}\beta_{f,-q}\rangle_{i} $ and $ \langle\beta^{\dagger}_{f,q}\beta^{\dagger}_{f,-q}\rangle_{i} $. Both the latter two terms vanish without quench.

% Results
%*************************************************************************************************
\section{Transient dynamics}
\begin{figure}[htbp]
\begin{center}
\includegraphics[width=\columnwidth]{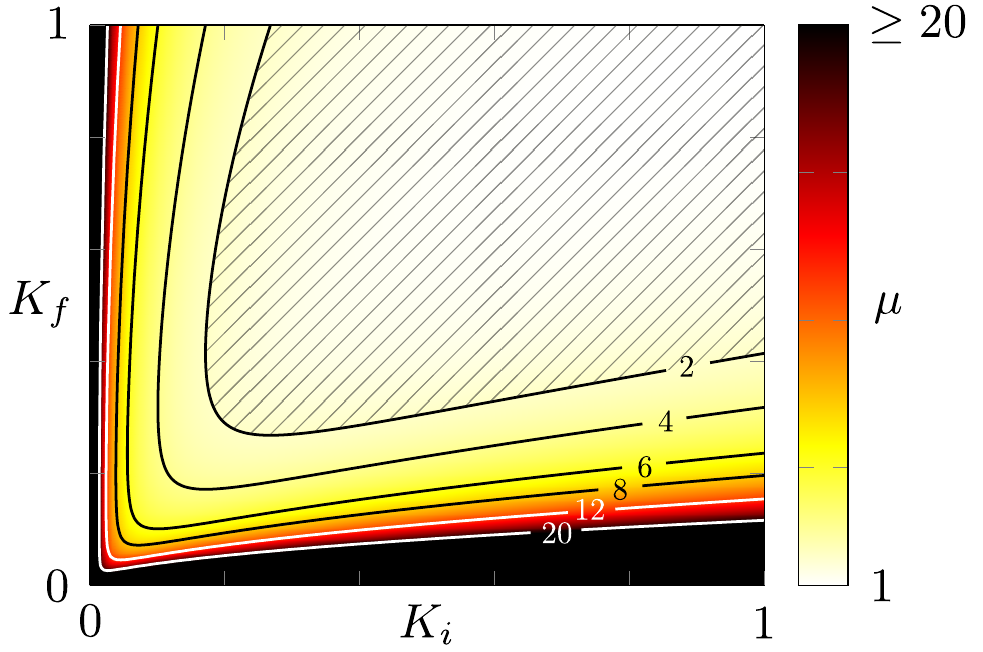}
\caption{(Color online) Contour plot of the non-universal coefficient $\mu$ as a function of $K_{i}$ and $K_{f}$. The dashed area is the region of the interaction parameters space $(K_i,K_f)$ where $1\leq\mu\leq 2$. Furthermore, the latter roughly corresponds to the region where the energy injected during the quench, $ \Omega_\nu $, is smaller than the Fermi energy. }
\label{fig:fig0}
\end{center}
\end{figure}

The behavior of the tunneling current of Eq.~(\ref{eq:current}) close to its steady state value $ I(\infty) $ is given by the following asymptotic expansion (see Appendix~\ref{appendix-transient} for details)
\begin{equation}
\label{eq:currentapprox}
I(t)\approx I(\infty)+\frac{I_1}{t^2}+\Delta I(t),
\end{equation}
where the limit $t_0\to 0$, justified for $t\gg t_0$, has been performed, $ I(\infty) $ is reported in the subsequent Section (note that $ I(\infty) $ depends on $ V $, see Eq.~\eqref{eq:currentasymp}) and
\begin{align}
I_1&=\frac{\eta}{4e^2}\frac{\partial^2}{\partial V^2}I(\infty)\, ,\\
\Delta I(t)&=I_0\left(\frac{\eta}{4}-1\right)\frac{\cos\left[\frac{\pi}{2}(\nu_{-}-\nu_{+})\right]}{eV\mathcal{K}^{\mu}t^{\mu+1}}\cos(eVt)\, .
\end{align}
Here, $ \mu=\nu_{-}+\nu_{+}\geq1 $, see Fig.~\ref{fig:fig0}, and thus the oscillating term is subdominant compared to the universal power-law. For $K_{f}=1$, i.e. quenching into a noninteracting system, one finds $I_{1}=0$ (since in this case $ \eta=0 $, see Eq.~\eqref{eq:eta}) and thus the tunneling current is predicted to show a non-universal decay. Our result differs from the ones found in Refs.~\onlinecite{Perfetto:2010,Schiro:2015} where a {\em non-universal} power-law was found. Although the universal power-law is a consequence of the peculiar intrinsic dynamics of a quenched LL -- see Appendix~\ref{appendix} -- the presence of an external probe is indeed essential for its observability~\cite{foot2}.
\noindent The universal scaling found here is also in contrast with the more standard, {\em non-universal} scaling of the tunneling current found in the absence of quench ($ K_i=K_f=K $)
\begin{equation}
I_{\mathrm{NQ}}(t)\approx I_{\mathrm{NQ}}(\infty)+I_2\frac{\cos(eVt)}{t^{\mu+1}}\, ,
\end{equation}
with $\mu$ evaluated here for $K_{i}=K_{f}=K$ and $ I_2=I_0\cos(\pi\nu_{-}/2)/(eV\mathcal{K}^{\mu}) $. 
\begin{figure}[htbp]
\begin{center}
\includegraphics[width=\columnwidth]{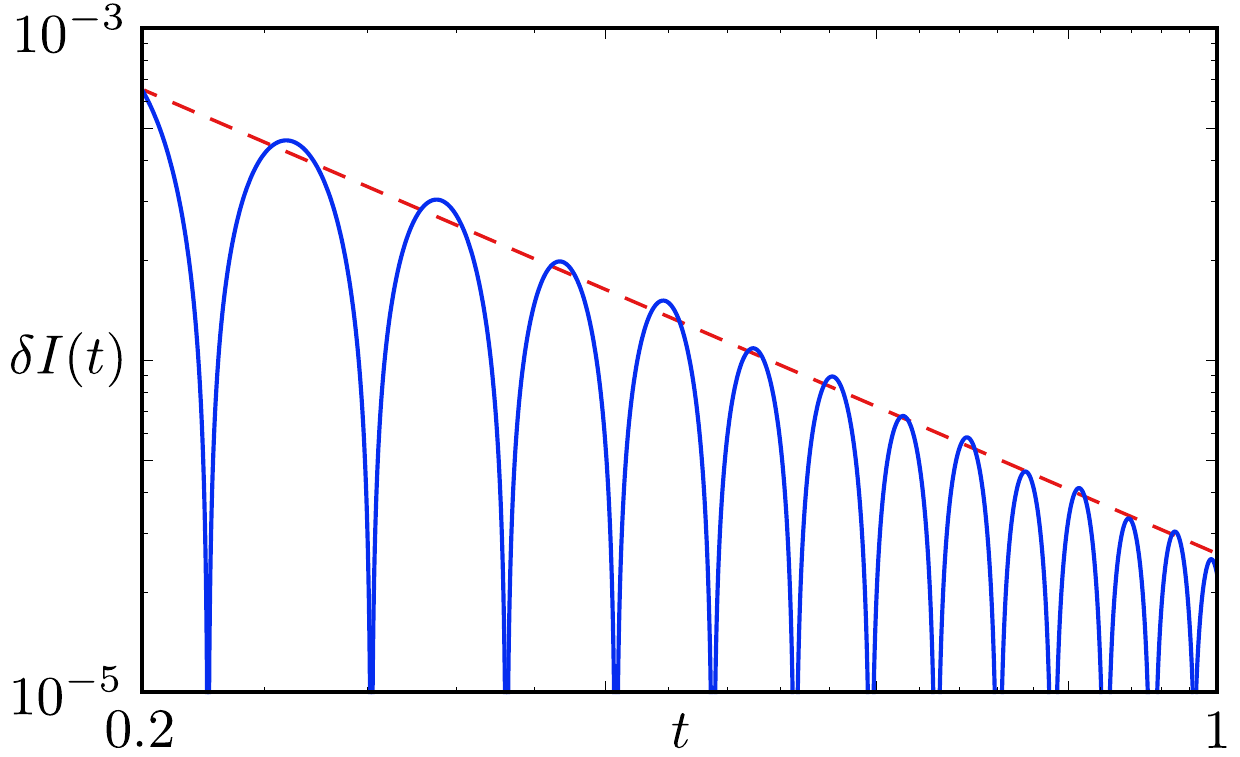}
\caption{(Color online) Solid: plot of $\delta I(t)$ as a function of $t$ (units $ 10^2\,\varepsilon_F^{-1} $) for $V=0.5$ (units $e^{-1}\varepsilon_F$) and $K_{i}=0.9$, $K_{f}=0.6$.  Here, $ \varepsilon_F $ is the Fermi energy of system and we have set $ \tau_0^{-1}=5\varepsilon_F $. The analytical scaling law $\propto t^{-2}$ is shown as a dashed guide to the eye. }
\label{fig:fig1}
\end{center}
\end{figure}
\begin{figure}[htbp]
\begin{center}
\includegraphics[width=\columnwidth]{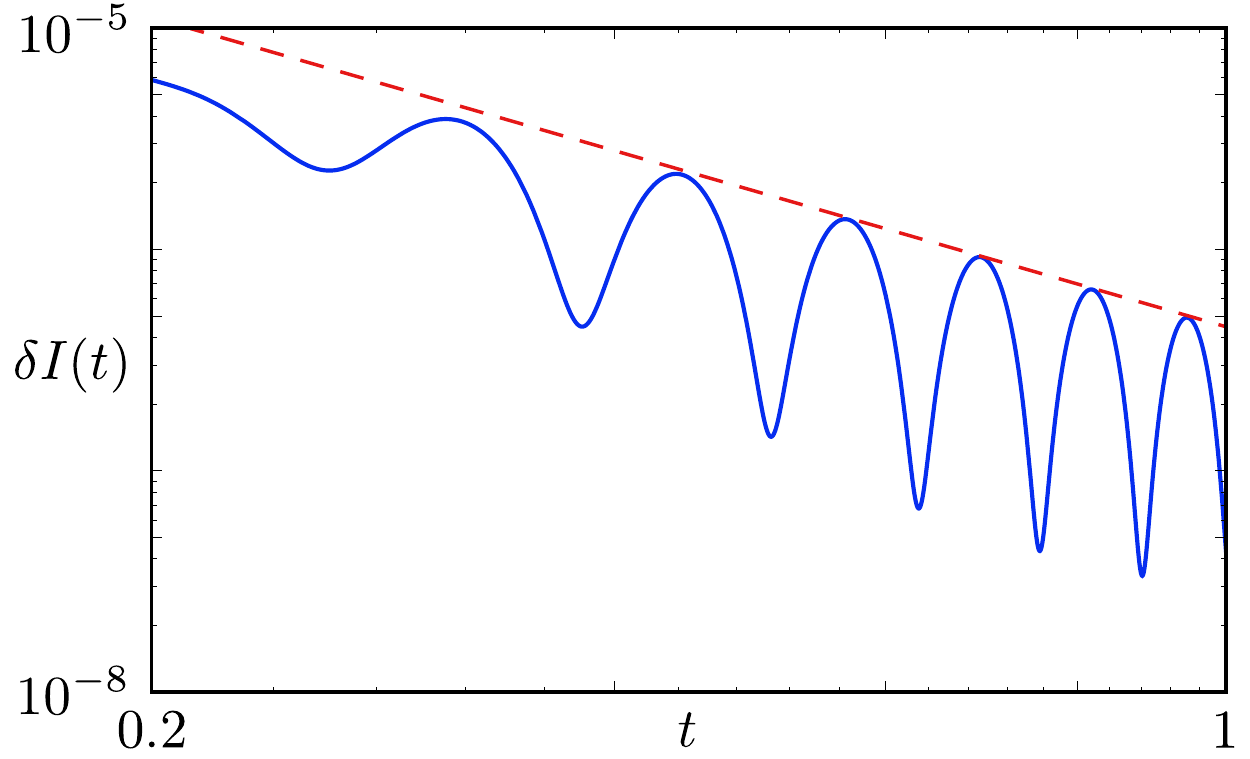}
\caption{(Color online) As in Fig.~\ref{fig:fig1} but for $K_{i}=0.4$, $K_{f}=0.9$.}
\label{fig:fig3}
\end{center}
\end{figure}
A better insight in the transient behavior of the tunneling current is obtained by looking at the relative quantity $\delta I(t)=[I(t)-I(\infty)]/I(\infty)$, shown in Figs.~\ref{fig:fig1}-\ref{fig:fig3} for different set of quench parameters. In particular, Fig.~\ref{fig:fig1} shows the case when interactions in the LL are {\em increased} (with $K_{f}<K_{i}$), while in Fig.~\ref{fig:fig3} the opposite case is displayed. In order to verify the universal power-law behavior predicted in Eq.~\eqref{eq:currentapprox}, in both Figs.~\ref{fig:fig1}-\ref{fig:fig3} the dashed line reproduces the power-law $\propto t^{-2}$ scaling. As one can see, the agreement between the envelope of $ \delta I(t) $ and the latter is very good in both cases. 

Since the universal time-scaling found in the transient regime arises from the term $ \mathcal{U}(t_1,t_2) $ in the correlator of Eq.~\eqref{eq:fb}, it can be understood by considering the propagation of entangled compound excitations~\cite{Calabrese:2006,Bernier:2014,Cazalilla:2016}. Indeed, as pointed out by Calabrese and Cardy~\cite{Calabrese:2006}, the pre-quench initial state acts as a source of these peculiar excitations. At any given time $ t>0 $, entangled pairs of excitations are created ($ \beta_{f,q}^\dagger \beta_{f,-q}^\dagger$) and annihilated ($ \beta_{f,q}\beta_{f,-q}$).
These entangled compound excitations propagate freely-like in opposite direction in the system and lead to a universal contribution to the tunneling current, independent from the interaction parameters. \\
Transport properties thus allows to probe the propagation of these entangled compound excitations in the LL. 
\section{Steady-state}
\begin{figure}[htbp]
\begin{center}
\includegraphics[width=\columnwidth]{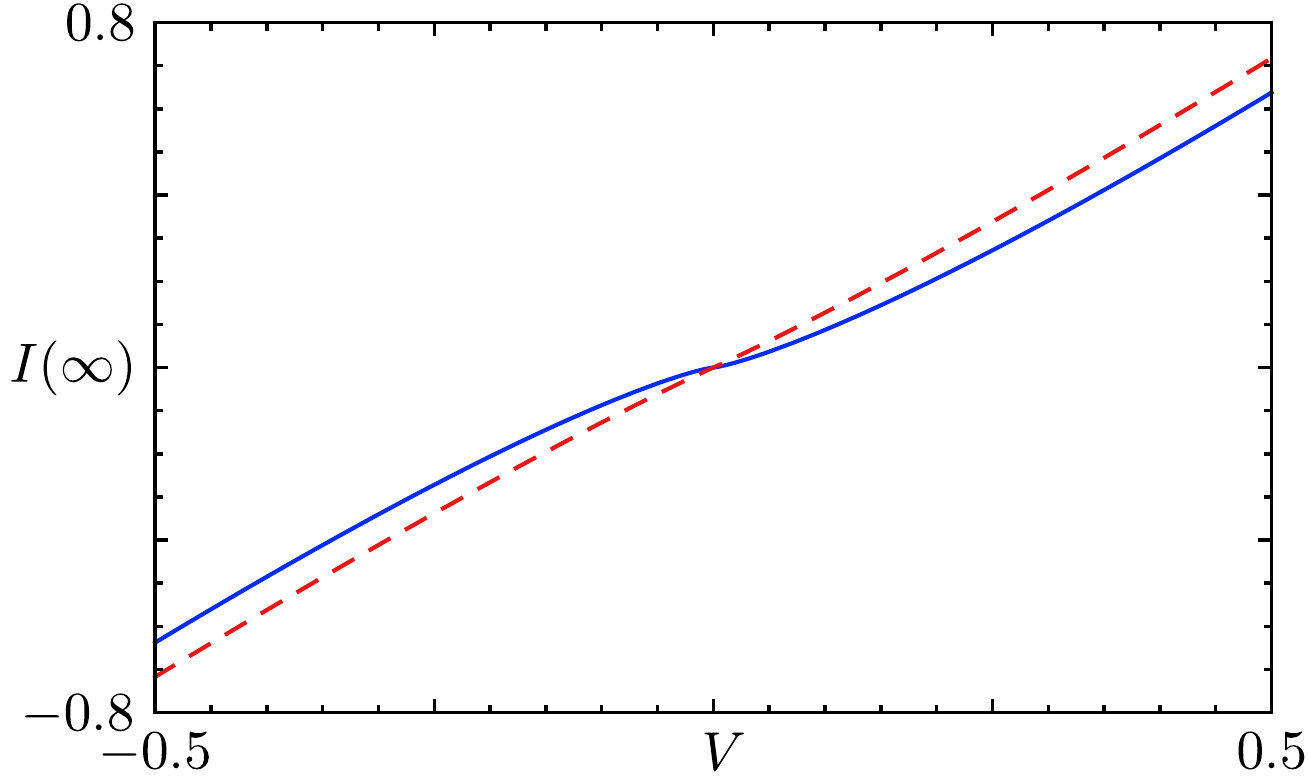}
\caption{(Color online) Plot of the asymptotic tunneling current $I(\infty)$ (units $e\mathcal{D}|\mathcal{M}|^2\varepsilon_F/(\pi v_F)$) as a function of the applied bias $V$ (units $e^{-1}\varepsilon_F$) for (solid) $K_{i}=K_{f}=0.5$ and (dashed) $K_{i}=0.9$, $K_{f}=0.5$. Here, $ \varepsilon_F $ and $ v_F $ are the Fermi energy and the Fermi velocity of the system respectively and we have set $ \tau_0^{-1}=5\varepsilon_F $. }
\label{fig:fig8}
\end{center}
\end{figure}

In the long-time limit the asymptotic tunneling current can be conveniently rewritten as
\begin{equation}
I(\infty)=2e\mathcal{D}|\mathcal{M}|^{2}\int_{-eV}^{eV}\mathrm{d}E\,\rho_T(E)\,,
\end{equation}
with the tunneling density of states (TDoS)
\begin{widetext}
\begin{equation}
\rho_T(E)=\frac{1}{\alpha\mathcal{K}}e^{-|E|\mathcal{K}^{-1}}(|E|\mathcal{K}^{-1})^{\mu-1}\left[\frac{U(\nu_{+},\mu,2|E|\mathcal{K}^{-1})}{\Gamma(\nu_{-})}\theta(-E)+\frac{U(\nu_{-},\mu,2|E|\mathcal{K}^{-1})}{\Gamma(\nu_{+})}\theta(E)\right]\, ,
\end{equation}
\end{widetext}
 where $ U(a,b,z) $ is the Tricomi confluent hypergeometric function. For $ V\rightarrow0 $ one obtains the expansion
 \begin{equation}
 \label{eq:currentasymp}
 I(\infty)\approx I^{\infty}_{1}V+\Delta I(\infty)\, ,
 \end{equation}
 with
 \begin{align}
 I^{\infty}_{1}&=\frac{I_0e}{\mathcal{K}}\frac{2^{1-\mu}\Gamma(\mu-1)}{\Gamma(\nu_{-})\Gamma(\nu_{+})}\,,\\
 \Delta I(\infty)&=I_0\frac{\Gamma(1\!-\!\mu)}{2\mu}\!\left[\sin(\pi\nu_{-})\!+\!\sin(\pi\nu_{+})\right]\!\frac{|V|}{V}\!\left(\!\frac{e|V|}{\mathcal{K}}\!\right)^{\!\mu}.
 \end{align}
Thus, since $ \mu>1 $, for small bias the asymptotic tunneling current is ohmic, i.e. linear in $ V $. Note that, in the absence of quench $ I_1^{\infty}=0 $ and the usual non-universal power-law voltage-scaling is recovered. Figure~\ref{fig:fig8} shows the behavior of $ I(\infty) $ as a function of the applied bias voltage for a non-quenched LL and for a quenched LL. The former case is represented as a solid line and is characterized by a tunneling current vanishing with a non-universal power-law for $ V\rightarrow0 $. In sharp contrast, in the presence of an interaction quench (dashed line) the tunneling current vanishes linearly as $ V\rightarrow0$, as expected from Eq.~\eqref{eq:currentasymp}. This behavior, already noted for the TDoS in Refs.~\cite{Kennes:2013,Kennes:2014} for quenches with $K_{i}=1$, is reminiscent of the case of a finite temperature LL~\cite{Kane:1992} and can be explained in terms of an effective quench-activated ``heating" of the LL. Indeed, it emerges from the fact that $ \langle\beta^{\dagger}_{f,q}\beta_{f,q}\rangle_{i}\neq0 $, similarly to what occurs in a thermally excited LL. This effective heating of the LL also has consequences on the asymptotic differential conductance, in which the zero-bias suppression found for a non-quenched zero-temperature LL~\cite{Giamarchi:2004,Voit:1995,vonDelft:1998} disappears, as can be directly verified from Eq.~\eqref{eq:currentasymp}.
\noindent A further evidence of the effective heating effect can be better understood by studying the LL asymptotic absorbed power as a function of the applied bias, which for a quenched LL becomes negative at low bias, implying that energy flows from the LL to the zero-temperature reservoir~\cite{Unp}. 

\section{Conclusions}
In this Letter we have demonstrated a universal behavior in the long-time dynamics following an interaction quench for a LL tunnel-coupled to a fermionic reservoir.
In contrast to what one would expect in the weak tunneling regime, where the dynamics following the quench should be dominated by the low-frequency modes of the LL and thus should be non-universal, we find universal contributions to the tunneling current. We explain this peculiar behavior in terms of free-like counter-propagating entangled pairs of excitations. We expect that the addition of higher order terms in the tunnel coupling would result in subleading contributions to the tunneling current and would not affect its universal behavior. \\
\noindent One could probe the predicted behavior using a quantum point contact (QPC) imprinted by optical means at the center of a trapped cloud of fermionic lithium atoms~\cite{Krinner:2015}. The QPC is subject to a controlled bias via its connection to particle reservoirs with different particle numbers, yielding a quasi-steady state current and giving direct access to its transport coefficients. The interaction strength, instead, can be adjusted and varied by a magnetic field as done in recent experiments with cold atoms~\cite{Krinner:2015,Husmann:2015,Krinner:2015b}.

\begin{acknowledgements}
The authors acknowledge the financial support of project MIUR-FIRB-2012-HybridNanoDev (Grant No. RBFR1236VV).
\end{acknowledgements}

\appendix
\section{Evaluation of the tunneling current }\label{appendix}

In this Appendix we will outline the derivation of the expression for the tunneling current $I(t)$ of Eq.~\eqref{eq:current}. To the lowest pertubative order in the tunnel coupling, i.e. in the sequential tunneling regime, the instantaneous tunneling current flowing from the reservoir to the LL is 
\begin{equation}
	\label{supp:current}
	I(t)=e\left[\Gamma^{+}(t)-\Gamma^{-}(t)\right]\,,
\end{equation} 
where $ \Gamma^{+}(t) $ is the tunnel-in rate and $ \Gamma^{-}(t) $ the tunnel-out rate. In order to obtain the rates $ \Gamma^{\pm}(t) $, we start evaluating the generic tunneling rate~\cite{WeissBook,Ziani:2012} from the initial state $|I\rangle=|N_{\mathrm{R,I}}\rangle|N_{\mathrm{LL,I}}\rangle$ to the final state $|F\rangle=|N_{\mathrm{R,F}}\rangle|N_{\mathrm{LL,F}}\rangle$, where  $N_{\mathrm{R,I}} $ ($N_{\mathrm{R,F}} $) and $ N_{\mathrm{LL,I}} $ ($ N_{\mathrm{LL,F}} $) are the number of particles in the initial (final) state in the reservoir and in the LL respectively,
\begin{equation}
	\Gamma_{\mathrm{I}\to \mathrm{F}}(t)=\frac{\partial}{\partial t}{\mathcal{P}_{\mathrm{I}\to \mathrm{F}}(t)}\,,
\end{equation}
where
\begin{equation}
	\label{eq:supp:prob}
	{\mathcal{P}}_{\mathrm{I}\to \mathrm{F}}(t)=\mathrm{Tr}_{\mathrm{R,LL}}\left\{\langle F|\rho_{I}(t)|F\rangle\right\},
\end{equation}
with $ \mathrm{Tr}_{\mathrm{R,LL}} $ the trace over the reservoir ($ \mathrm{R} $) and LL bosonic excitations ($ \mathrm{LL} $). Here $ \rho_I(t) $ is the density matrix of the total system in the interaction picture with respect to tunneling Hamiltonian of Eq.~\eqref{eq:H_T}, 
\begin{align}
	H_{\mathrm{T}}(t)&=\mathcal{M}\theta(t-t_0)\sum_{r=\pm}\psi_{r}^{\dagger}(x_0)\psi_{\mathrm{R}}(z_{\mathrm{R}})+\mathrm{H.c.}\nonumber \\
	&\equiv H_{\mathrm{T}}^{+}(t)+H_{\mathrm{T}}^{-}(t)\, .
\end{align} 
Our first task is to obtain an expression for the time evolution of $\rho_{I}(t)$ for $t>0^{+}$ (i.e. after the quench), provided the system is in thermal equilibrium immediately before the quench, i.e. it is described by the equilibrium density matrix
\begin{equation}
	\rho(0^-)=\rho_{\mathrm{R}}\rho_{\mathrm{LL}}|N_{\mathrm{R,I}}\rangle\langle N_{\mathrm{R,I}}||N_{\mathrm{LL,I}}\rangle\langle N_{\mathrm{LL,I}}|,
\end{equation}
with $ \rho_{\mathrm{R}} $ ($ \rho_{\mathrm{LL}} $) the equilibrium density matrix for the reservoir (LL). In the zero-temperature limit $ \rho(0^-) $ reduces to
\begin{equation}
	\rho(0^{-})=\prod_{i=\mathrm{R,LL}}|\Omega_{i}\rangle\langle\Omega_{i}||N_{i,\mathrm{I}}\rangle\langle N_{i,\mathrm{I}}|\, ,
\end{equation}
where $|\Omega_{i}\rangle$ is the ground state for the quasiparticles ($i=\mathrm{R}$) or collective ($i=\mathrm{LL}$) excitations of reservoir and LL respectively. In the interaction picture the time evolution of the density matrix is given by
\begin{equation}
	\rho_{I}(t)=\mathcal{T}\left[e^{-i\int_{0}^{t}{\mathrm{d}}t'\ H_{\mathrm{T},I}(t')}\right]\rho(0^{-})\widetilde{\mathcal{T}}\left[e^{i\int_{0}^{t}{\mathrm{d}}t'\ H_{\mathrm{T},I}(t')}\right]\, ,
\end{equation}
where $\mathcal{T}$ and $\widetilde{\mathcal{T}}$ denote time-ordering and anti-time-ordering operators respectively. We expand the time evolution operators to lowest order in $H_{\mathrm{T},I}(t')$ and plug the corresponding expression into Eq.~\eqref{eq:supp:prob}. This results in the following selection rules for $N_{\mathrm{R,F}}$ and $N_{\mathrm{LL,F}}$: 
$N_{\mathrm{R,F}}=N_{\mathrm{R,I}}\mp 1$ and $N_{\mathrm{LL,F}}=N_{\mathrm{LL,I}}\pm 1$, which describe tunnel-in (with rate $ \Gamma^{+}(t) $) and tunnel-out (with rate $ \Gamma^{-}(t) $) events.
For the sake of brevity, we outline the procedure for tunnel-in events and thus choose $N_{\mathrm{LL,F}}=N_{\mathrm{LL,I}}+1$ and accordingly $N_{\mathrm{R,F}}=N_{\mathrm{R,I}}-1$. By virtue of this one gets
\begin{align}
	\mathcal{P}_{\mathrm{I}\to \mathrm{F}}^{+}(t)&=\mathrm{Tr}_{\mathrm{R,LL}}\bigg\{\int_{t_0}^{t}\mathrm{d}t'\int_{t_0}^{t}\mathrm{d}t''\nonumber\\
	&\times\langle F|H_{\mathrm{T}}^{+}(t')\rho_{I}(0^{-})H_{\mathrm{T}}^{-}(t'')|F\rangle\bigg\}\, ,
\end{align}
which can be rewritten as
\begin{equation}
	\label{eq:P1}
	\mathcal{P}_{\mathrm{I}\to \mathrm{F}}^{+}(t)=|\mathcal{M}|^2\sum_{r=\pm}\int_{t_0}^{t}\mathrm{d}t'\int_{t_0}^{t}\mathrm{d}t''\ f_{b,r}^{+}(t',t'')f_{\mathrm{R}}^{+}(t'-t'')\, ,
\end{equation}
with
\begin{align}
	f_{b,r}^{+}(t',t'')&=\mathrm{Tr}_{\mathrm{LL}}\left\{\langle N_{\mathrm{LL,I}}|\Psi_{r,I}(x_0,t'')\Psi_{r,I}^{\dagger}(x_0,t')\right.\nonumber\\ &\left.\times\frac{1}{Z_{\mathrm{LL}}}e^{-\beta H_{0,\mathrm{LL}}}|N_{\mathrm{LL,I}}\rangle\right\}
\end{align}
and
\begin{align}
	f_{\mathrm{R}}^{+}(t'-t'')&=\mathrm{Tr}_{\mathrm{R}}\left\{\langle N_{\mathrm{R,I}}|\Psi_{\mathrm{R},I}^{\dagger}(z_{\mathrm{R}},0)\Psi_{\mathrm{R},I}(z_{\mathrm{R}},t'-t'')\right.\nonumber\\
	&\left.\times\frac{1}{Z_{\mathrm{R}}}e^{-\beta H_{0,\mathrm{R}}}|N_{\mathrm{R,I}}\rangle\right\}\, .	
\end{align}
To proceed, the integrations in Eq.~\eqref{eq:P1} are shifted by $t_0$ and integration domain is split:
\begin{equation}
	\label{eq:P2}
	\mathcal{P}_{\mathrm{I}\to \mathrm{F}}^{+}(t)=\int_{0}^{t-t_0}\mathrm{d}t'\ \left[F_1^{+}(t')+F_2^{+}(t')\right]\, ,
\end{equation}
with
\begin{align}
	F_1^{+}(t')&\!=\!|\mathcal{M}|^2\!\!\sum_{r=\pm}\!\int_{0}^{t'}\mathrm{d}t''\, f_{b,r}^{+}(t'+t_0,t''+t_0)f_{\mathrm{R}}^{+}(t'-t'')\, ,\\
	F_2^{+}(t')&\!=\!|\mathcal{M}|^2\!\!\sum_{r=\pm}\!\int_{0}^{t'}\mathrm{d}t''\, f_{b,r}^{+}(t''+t_0,t'+t_0)f_{\mathrm{R}}^{+}(t''-t')\, .
\end{align}
The tunnel-in tunneling rate is thus given by
\begin{equation}
	\Gamma^{+}(t)=2|\mathcal{M}|^2\sum_{r=\pm}\int_{0}^{t-t_0}\mathrm{d}t'\ \mathrm{Re}\left[f_{\mathrm{R}}^{+}(t')f_{b,r}^{+}(t,t-t')\right]\, .	
\end{equation}
The evaluation of correlation functions proceeds as follows. Using Eq.~\eqref{eq:psiT}, the reservoir contribution is
\begin{equation}
	f_{\mathrm{R}}^{+}(t)=\sum_{\mathbf{k}}|\Psi_{\mathbf{k}}(z_{\mathrm{R}})|^2f(\epsilon_{\mathbf{k}}-eV)e^{-i\epsilon_{\mathbf{k}}t}\, ,
\end{equation}
where $f(E)$ is the Fermi function and $ V $ the bias between the reservoir and the LL. Assuming a weak dependence of $\Psi_{\mathbf{k}}(z_{\mathrm{R}})$ on $z_{\mathrm{R}}$ and converting the sum over momenta to an integration on the energy, we finally obtain
\begin{equation}
	f_{\mathrm{R}}^{+}(t)=\mathcal{D}\int_{-E_{\mathrm{F}}}^{\infty}\mathrm{d}E\ f(E-eV)e^{-iEt}\, ,
\end{equation}
where $E_{\mathrm{F}}$ is the Fermi energy and $ \mathcal{D} $ the density of states of the reservoir, respectively. Concerning the LL contribution, from Eq.~\eqref{eq:psir}, we find
\begin{align}
	f_{b,r}^{+}(t_1,t_2)&=\frac{1}{2\pi\alpha}e^{-\frac{1}{2}\langle\left[\phi_{r,I}(x_0,t_2)-\phi_{r,I}(x_0,t_1)\right]^2\rangle_i}\nonumber\\ & \times e^{\frac{1}{2}\left[\phi_{r,I}(x_0,t_2),\phi_{r,I}(x_0,t_1)\right]}\, ,	
\end{align}
where $\langle\ldots\rangle_i$ denotes the average on the thermal distribution of the bosonic eigenstates for $t<0^{-}$. Upon expressing the fields $\phi_{r,I}(x,t)$ of Eq.~\eqref{eq:phir} in terms of the operators $\beta_{i,q}$ and $\beta_{i,q}^{\dagger}$~\cite{Cazalilla:2006,Iucci:2009}, see Eq.~\eqref{eq:betaf}, the correlator in the zero-temperature limit can be evaluated to
\begin{equation}
	f_{b,r}^{+}(t_1,t_2)=\frac{1}{2\pi\alpha}e^{-S(t_1,t_2)}\, ,	
\end{equation}
with
\begin{widetext}
\begin{equation}
	S(t_1,t_2)=\frac{2\pi}{L}\sum_{q>0}\frac{e^{-\alpha q}}{q}\left\{\gamma+\eta\left[\cos(2v_{f}qt_2)+\cos(2v_{f}qt_1)-2\cos(2v_{f}q(t_1+t_2))\right]-\nu_{-} e^{iv_{f}q(t_1-t_2)}-\nu_{+} e^{-iv_{f}q(t_1-t_2)}\right\}\, ,\label{eq:Sum}
\end{equation}
\end{widetext}
where
\begin{subequations}
	\begin{align}
		\gamma&=\frac{(1+K_{f}^2)(K_{f}^2+K_{i}^2)}{4K_{f}^2K_{i}}\, ,\\
		\eta&=\frac{(1-K_{f}^2)(K_{f}^2-K_{i}^2)}{8K_{f}^2K_{i}}\, ,\label{supp:eq:c2}\\
		\nu_{\pm}&=\frac{(1+K_{f}^2)(K_{f}\mp K_{i})^2}{8K_{f}^2K_{i}}\, .\label{supp:eq:nu}
	\end{align}
\end{subequations}
Note that the result is independent of $r$. The sums in Eq.~\eqref{eq:Sum} can be evaluated analytically and, performing the limits $\alpha/L\to 0$ and $v_{f}t/L\to 0$, one gets
\begin{align}
	\label{eq:S}
	f_{b}(t_1,t_2)&=\left[\frac{1}{1-i\mathcal{K}(t_1-t_2)}\right]^{\nu_{-}}\left[\frac{1}{1+i\mathcal{K}(t_1-t_2)}\right]^{\nu_{+}}\nonumber\\
	&\times\left\{\frac{(1+4\mathcal{K}^2t_2^2)(1+4\mathcal{K}^2t_1^2)}{\left[1+\mathcal{K}^2(t_1+t_2)^2\right]^2}\right\}^{\eta/2}\, ,
\end{align}
where $\mathcal{K}=(\tau_0 K_{f})^{-1}$ and we have introduced the notation $(2\pi\alpha)^{-1}f_{b}(t_1,t_2)\equiv f_{b,R}(t_1,t_2)=f_{b,L}(t_1,t_2)$. Thus, one gets
\begin{align}
	\label{supp:Gamma+}
	\Gamma^{+}(t)&=\frac{4\mathcal{D}|\mathcal{M}|^2}{2\pi\alpha}\int_{-\infty}^{\infty}\mathrm{d}E\, f(E-eV)\nonumber\\
	&\times\int_{0}^{t-t_0}\mathrm{d}t'\ \mathrm{Re}\left[e^{-iEt'}f_{b}(t,t-t')\right]\, ,
\end{align}
having assumed $E_{\mathrm{F}}\gg|eV|$. With an analogous procedure we obtain the tunnel-out rate, which is given by
\begin{align}
	\label{supp:Gamma-}
	\Gamma^{-}(t)&=\frac{4\mathcal{D}|\mathcal{M}|^2}{2\pi\alpha}\int_{-\infty}^{\infty}\mathrm{d}E\,\left[1-f(E-eV)\right]\nonumber\\
	&\times\int_{0}^{t-t_0}\mathrm{d}t'\ \mathrm{Re}\left[e^{iEt'}f_{b}(t,t-t')\right]\, .
\end{align}
\subsection{Steady-state regime}
The steady state regime is obtained in the limit $ t\rightarrow\infty $. For the tunnel-in rate we have
\begin{equation}
	\Gamma^{+}(\infty)=2\mathcal{D}|\mathcal{M}|^2\int_{-\infty}^{\infty}\mathrm{d}E\, f(E-eV)\rho_T(E),
\end{equation}  
where we have introduced the tunneling density of states (TDoS)
\begin{equation}
	\rho_T(E)\equiv\frac{1}{2\pi\alpha}\int_{-\infty}^{\infty}\mathrm{d}t'\,e^{-iEt'}f_b^\infty(t')\,,
\end{equation}
with
\begin{equation}
	f_b^\infty(t')=\left[\frac{1}{1-i\mathcal{K}t'}\right]^{\nu_{-}}\left[\frac{1}{1+i\mathcal{K}t'}\right]^{\nu_{+}}\, .
\end{equation}
Analytically one obtains
\begin{widetext}
\begin{equation}
	\rho_T(E)=\frac{1}{\alpha\mathcal{K}}e^{-|E|\mathcal{K}^{-1}}(|E|\mathcal{K}^{-1})^{\nu_{-}+\nu_{+}-1}\left[\frac{U(\nu_{+},\nu_{-}+\nu_{+},2|E|\mathcal{K}^{-1})}{\Gamma(\nu_{-})}\theta(-E)+\frac{U(\nu_{-},\nu_{-}+\nu_{+},2|E|\mathcal{K}^{-1})}{\Gamma(\nu_{+})}\theta(E)\right],
\end{equation} 
\end{widetext}
with $ U(a,b,z) $ the Tricomi confluent hypergeometric function. From Eqs.~\eqref{supp:current},\eqref{supp:Gamma+} and \eqref{supp:Gamma-}, we thus obtain the steady state tunneling current 
\begin{equation}
	\label{supp:eq:Iinfty}
	I(\infty)=2e\mathcal{D}|\mathcal{M}|^{2}\int_{-eV}^{eV}\mathrm{d}E\,\rho_T(E)\,.
\end{equation}
Furthermore, for small bias, i.e. for $ eV\mathcal{K}^{-1}\ll 1 $, the following expansion holds
\begin{widetext}
\begin{equation}
	I(\infty)\approx I_0\left\{\frac{2^{1-\nu_{-}-\nu_{+}}\pi\Gamma(\nu_{-}+\nu_{+}-1)}{\Gamma(\nu_{-})\Gamma(\nu_{+})}\frac{eV}{\mathcal{K}} +\frac{1}{2}\frac{\Gamma(1-\nu_{-}-\nu_{+})}{\nu_{-}+\nu_{+}}\left[\sin(\pi \nu_{-})+\sin(\pi \nu_{+})\right]\frac{|V|}{V}\left(\frac{e|V|}{\mathcal{K}}\right)^{\nu_{-}+\nu_{+}}\right\}\, ,
\end{equation}
\end{widetext}
where $I_0=4e\mathcal{D}|\mathcal{M}|^2/(\pi\alpha)$. Since $ \nu_{-}+\nu_{+}\geq1 $, the second term is always subleading and the steady state current is linear in the small bias. However, in the absence of quench, i.e. when $ K_i=K_f=K $, $ \nu_{+}=0 $ (see Eq.~\eqref{supp:eq:nu} ) and, since $ [\Gamma(x)]^{-1}\approx x $ for $ x\rightarrow0 $, the usual power-law behavior of a non-quenched LL is recovered. 
\subsection{Transient regime}\label{appendix-transient}
In order to study how the steady state regime is approached we start again from Eq.~\eqref{supp:current}, that after the integration over energies reads
\begin{equation}
	\label{supp:current2}
	I(\bar{t})=I_0\int_{0}^{\bar{t}}\mathrm{d}\bar{\tau}\,\mathrm{Re}\left[\frac{\sin(\bar{V}\bar{\tau})}{\bar{\tau}}\bar{f}_b(\bar{t},\bar{t}-\bar{\tau})\right]\, ,
\end{equation}
where for future convenience we have switched to the dimensionless variables $\bar{t}=t/\tau_0$, $ \bar{\tau}=t'/\tau_0 $, $\bar{V}=eV\tau_0$ and the limit $t_0\to 0$, justified for $t\gg t_0$, has been performed~\cite{foot1}. Here 
\begin{align}
	\bar{f}_{b}(\bar{t},\bar{t}-\bar{\tau})&=\left(\frac{1}{1-iK_f^{-1}\bar{\tau}}\right)^{\nu_{-}}\left(\frac{1}{1+iK_f^{-1}\bar{\tau}}\right)^{\nu_{+}}\nonumber\\
	&\times\left\{\frac{(1+4K_f^{-2}\bar{t})\left[1+4K_f^{-2}(\bar{t}-\bar{\tau})^2\right]}{\left[1+K_f^{-2}(2\bar{t}-\bar{\tau})^2\right]^2}\right\}^{\eta/2}\, ,
\end{align}
Provided $ \bar{V}\bar{t}\gg1 $, inside the integral of Eq.~\eqref{supp:current2} this correlator can be approximated by
\begin{equation}
	f_{b}(\bar{t},\bar{t}-\bar{\tau})\!\approx\!\left(\frac{1}{1-iK_f^{-1}\bar{\tau}}\right)^{\nu_{-}}\!\left(\frac{1}{1+iK_f^{-1}\bar{\tau}}\right)^{\nu_{+}}\!\left(1-\frac{\eta \bar{\tau}^2}{4 \bar{t}^2}\right) .
\end{equation}
To proceed it is now convenient to rewrite the integral as 
\begin{equation} \int_{0}^{\bar{t}}\mathrm{d}\bar{\tau}=\left(\int_{0}^{\infty}-\int_{\bar{t}}^{\infty}\right)\mathrm{d}\bar{\tau}\,.
\end{equation}
In order to evaluate the first integral one can notice that 
\begin{align}
	&I_0\int_{0}^{\infty}\mathrm{d}\bar{\tau}\, \mathrm{Re} \left[\frac{\sin(\bar{V}\bar{\tau})}{\bar{\tau}}\left(\frac{1}{1-iK_f^{-1}\bar{\tau}}\right)^{\nu_{-}}\!\left(\frac{1}{1+iK_f^{-1}\bar{\tau}}\right)^{\nu_{+}}\right]\nonumber\\
	&=I(\infty)\, ,\\
	&I_0\int_{0}^{\infty}\!\!\mathrm{d}\bar{\tau}\, \mathrm{Re} \left[\bar{\tau}\sin(\bar{V}\bar{\tau})\left(\frac{1}{1-iK_f^{-1}\bar{\tau}}\right)^{\nu_{-}}\!\left(\frac{1}{1+iK_f^{-1}\bar{\tau}}\right)^{\nu_{+}}\right]\nonumber\\
	&=-\frac{\partial^2}{\partial\bar{V}^2}I(\infty)\, ,
\end{align}
with $ I(\infty) $ given in Eq.~\eqref{supp:eq:Iinfty}, while for the second integral one can safely expand for $ \bar{V}\bar{t}\gg1 $ (and $ \bar{t}\gg1 $), obtaining
\begin{widetext}
\begin{equation}
	\int_{1}^{\infty}\mathrm{d}x\,\frac{\sin(\bar{V}\bar{t}x)}{x}\left(\frac{1}{1-iK_f^{-1}x\bar{t}}\right)^{\nu_{-}}\left(\frac{1}{1+iK_f^{-1}x\bar{t}}\right)^{\nu_{+}}\left(1-\frac{\eta x}{4}\right)\approx\frac{\cos\left[\frac{\pi}{2}(\nu_{-}-\nu_{+})\right]}{\bar{V}\bar{t}}\left(\frac{K_f}{\bar{t}}\right)^{\nu_{-}+\nu_{+}}\left(1-\frac{\eta}{4}\right)\cos(\bar{V}\bar{t})\, ,
\end{equation}
\end{widetext}
where on the left-hand side we have performed the change of variable $ \bar{\tau}=x\bar{t} $.
Thus, we finally obtain
\begin{equation}
	I(\bar{t})=I(\infty)+\frac{I_1}{\bar{t}^2}+\Delta I(\bar{t}),
\end{equation}
where
\begin{subequations}
	\begin{align}
		I_1&=\frac{\eta}{4}\frac{\partial^2}{\partial\bar{V}^2}I(\infty)\, ,\\
		\Delta I(\bar{t})&=I_0\left(\frac{\eta}{4}-1\right)\frac{K_f^{\nu_{-}+\nu_{+}}\cos\left[\frac{\pi}{2}(\nu_{-}-\nu_{+})\right]}{\bar{V}\bar{t}^{\nu_{-}+\nu_{+}+1}}\cos(\bar{V}\bar{t})\,.
	\end{align}
\end{subequations}
Since $ \nu_{-}+\nu_{+}\geq1 $, $ \Delta I(\bar{t}) $ is a subleading contribution and thus the approach to the steady state is controlled by the term $ \propto \bar{t}^{-2} $.

 %*************************************************************************************************
 % Bibliography
 %*************************************************************************************************

%

\begin{thebibliography}{99}
\bibitem{foot2} In our work we have considered as external probe only a non-interacting fermionic reservoir. However, calculations similar to the ones performed in Appendix~\ref{appendix} show that the universal behavior also emerges with an interacting probe, such as another LL coupled to the first one through a quantum point contact.
\bibitem{foot1} Note that, since $ \tau_0 $ is the shortest-time cutoff of the theory, in the following we are interested in times $\bar{t}\gg1 $.
\end{thebibliography}
\end{document}